\documentclass[aps,prl, twocolumn, amsfonts, amssymb]{revtex4}
\usepackage{hyperref, graphicx, wrapfig, color}
\pdfoutput=1
\newcommand{\be}{\begin{equation}}
\newcommand{\ee}{\end{equation}}

\begin{document}
\title{Spontaneous emission in a silicon charge qubit}
\author{Khoi T. Nguyen}
\author{N. Tobias Jacobson}
\author{Michael P. Lilly}
\author{N. Bishop}
\author{E. Nielsen}
\author{J. Wendt}
\author{J. Dominguez}
\author{T. Pluym}
\author{Malcolm S. Carroll}
\email{mscarro@sandia.gov}
\affiliation{Sandia National Laboratories, Albuquerque, New Mexico 87185, USA}

\begin{abstract}
The interaction between a qubit and its environment provides a channel for energy relaxation which has an energy-dependent timescale governed by the specific coupling mechanism. 
We measure the rate of inelastic decay in a Si MOS double quantum dot (DQD) charge qubit through sensing the charge state's response to non-adiabatic driving of its excited state population.  
The charge distribution is sensed remotely in the weak measurement regime.  
We extract emission rates down to kHz frequencies by measuring the variation of the non-equilibrium charge occupancy as a function of amplitude and dwell times between non-adiabatic pulses. 
Our measurement of the energy-dependent relaxation rate provides a fingerprint of the relaxation mechanism, indicating that relaxation rates for this Si MOS DQD are consistent with coupling to deformation acoustic phonons.
\end{abstract}

\maketitle

\paragraph{Introduction-}
Spontaneous emission and absorption are fundamental processes of energy exchange between a two-level system and its environment. 
The emission and absorption rates for quantum dots (QD) depend on microscopic details such as the materials \cite{Fujisawa1998} and geometry\cite{Weig2004} of the particular experimental system, and are important factors for properties such as inelastic tunneling, excited state relaxation times \cite{Johnson2005} (e.g., $T_{1}$ for charge quantum bits), photon emission, and thermal equilibration times. 
Silicon quantum dots are of particular interest because of their promise for quantum information science, which is motivated by the ability to achieve very long electron spin coherence times with isotopic enrichment \cite{Witzel2010}. 
Interest has also intensified recently because of improvements in the design and fabrication of few-electron Si double quantum dots (DQD) that have achieved more ideal QD behavior\cite{Maune2012, Yang2012}.

Silicon has a non-polar crystal structure with weak electron-phonon piezoelectric coupling, and in many cases it is expected that acoustic phonons will dominate excited state relaxation. 
A recent examination of the relaxation time energy dependence, using a well-established technique called photon assisted tunneling, showed an energy dependence that was inconsistent with an acoustic phonon spectral density \cite{Wang2013}. 
There are few other direct quantitative measurements of silicon DQD excited charge state relaxation times and those, for example, use current through the DQD as a probe, limiting the measurable range to emission rates that directly produce detectable currents through the DQD \cite{Liu2008,Ogi2010}.  
Strong coupling to the DQD through the leads for current measurement, or even relatively strong remote sensing \cite{Pioro-Ladriere2005, Harbusch2010, Granger2012}, can furthermore perturb the DQD and its relaxation times, leading to a convolution of effects that obscures the underlying physics of the DQD's interaction with its environment.  
An improved and more complete measurement of relaxation times in Si DQDs is of significant interest to better understand this central property of Si quantum dot physics.  

In this Letter, we extract the energy-dependent spontaneous emission of a silicon double quantum dot using a method that can be generalized to other two-level systems.  
With a neighboring weakly-coupled charge sensor, we measure the steady-state charge distribution of a Si DQD charge qubit while subject to periodic diabatic pulsing of varying frequency and pulse amplitude. 
The variation of the charge distribution with these parameters depends strongly on the functional form of the energy-dependent spontaneous emission, providing a clear signature of the underlying microscopic physical mechanism responsible for the steady-state distribution. 
We extract the energy-dependent spontaneous emission rate in the Si DQD by fitting to the steady-state charge distribution, finding quantitative agreement with acoustic phonons as the dominant channel for energy relaxation.  
Estimates of the elastic tunnel coupling and spectral density can also be extracted with this approach.  This technique extends the range of measurable  spontaneous emission rates ($\sim$kHz in this case), which are slow relative to the charge decoherence time or measurable inelastic electron transport times.

\begin{figure}[h]
\begin{center}
\includegraphics[width=0.4\textwidth]{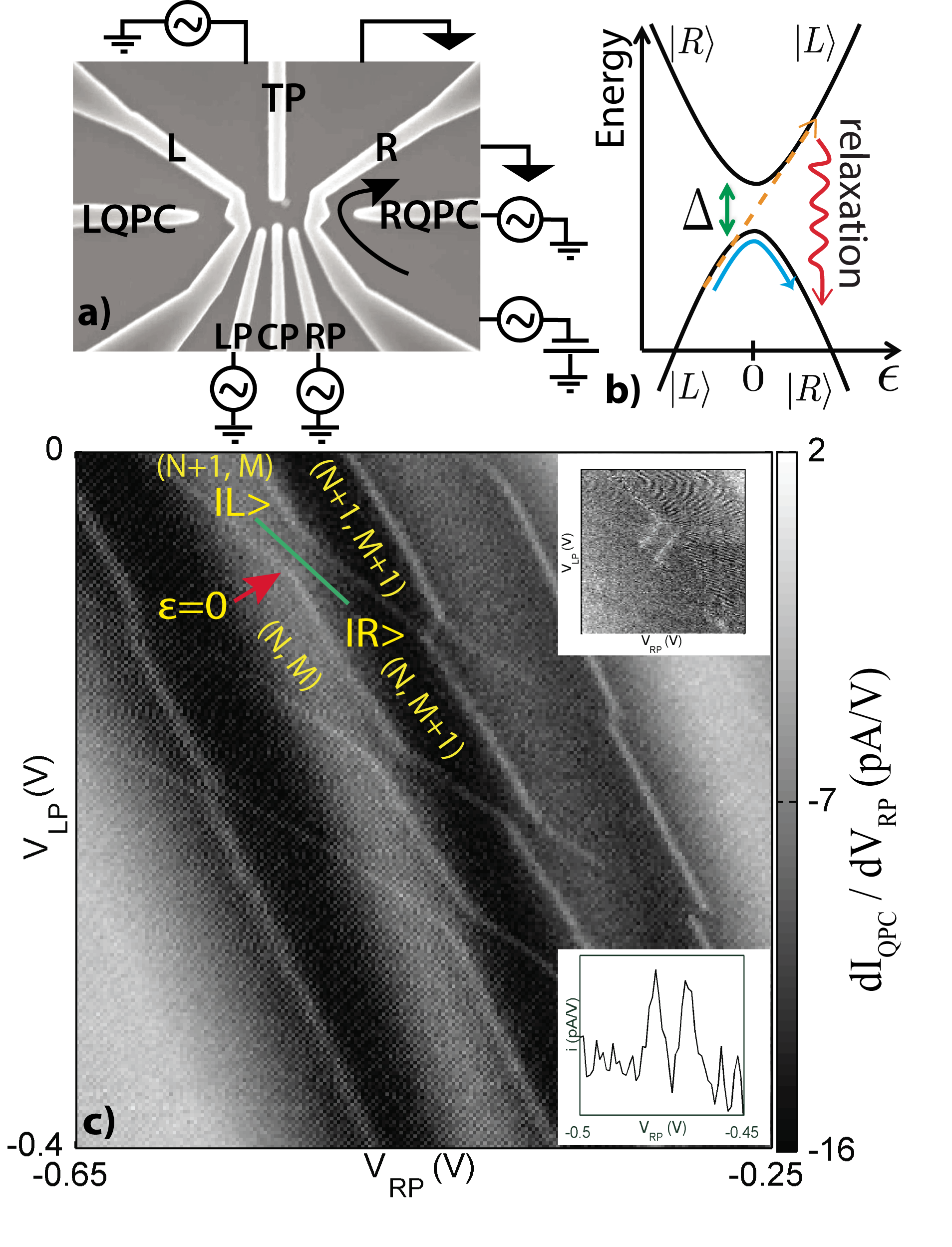}
\end{center}
\vspace{-20pt}
\caption{(Color online) 
(a) Scanning electron microscopy image of depletion gates, with measurement setup. An Al global gate above the depletion gates shown in the SEM is biased to 3.65 V, and the L, R, TP, CP, LQPC, and RQPC gates are biased to 
-0.2, -1.2, 0.05, -0.15, 0, and -7.8 V, respectively, 
(b) Schematic of ground and first excited state energies as a function of detuning.  Highlighted on the schematic are an adiabatic passage (blue) and diabatic passage (green dash) with relaxation (red), (c) Charge stability diagram of the double quantum dot obtained from the differential charge sensing measurement; $\vert L\rangle$ and $\vert R \rangle$ indicate the two charge sectors defining the charge qubit and the green line indicates the detuning direction across the charge transition line, 
(inset upper right) charge transition region between $V_{LP}$ and $V_{RP}$ when a 43 Hz square wave is applied, which doubles the transition line (dashed lines added to guide the eye), (inset lower right) analog derivative of the QPC current across the doubled charge transition line.}
\label{fig:DQD_SEM_Stability_Diagram}
\end{figure}

\paragraph{Experiment-}
We perform all measurements on the silicon DQD nanostructure shown in Fig. \ref{fig:DQD_SEM_Stability_Diagram}(a). The nominal fabrication steps for this device have been described previously \cite{Nordberg2009}.
The right constriction acts as a quantum point contact (QPC) charge sensor \cite{Field1993}. The conductance of the constriction is approximately $5 \ \mathrm{\mu S}$, and we estimate that the change in conductance is approximately 0.01\% when the electron occupation of the left dot changes by one.  We carry out the experiment in a dilution refrigerator at 25 mK base temperature, with a perpendicular B-field of 100 mT.  

Figure \ref{fig:DQD_SEM_Stability_Diagram}(b) shows a representative charge stability diagram of this system, measured with the QPC transconductance.  Charge sectors labeled $(N,M)$ correspond to regions in the $V_{LP}$-$V_{RP}$ plane for which, at equilibrium, the DQD has $N$ ($M$) charge quanta on the left (right) quantum dot.
The background slope in the charge sensor current is due to capacitive coupling between the charge sensor and neighboring gates, which slowly modulates the conductance.
The current through the QPC probes the charge distribution within the DQD.
In the differential QPC current plotted in Fig. \ref{fig:DQD_SEM_Stability_Diagram}(b), transitions between charge sectors appear as light lines on a dark background.
Here, we are interested in charge-conserving transitions of the form $(N+1, M) \leftrightarrow (N, M+1)$. To drive the system across this transition, we pulse between $V_{LP}$ and $V_{RP}$ along a diagonal sweep perpendicular to the transition boundary. 
The detuning voltage is the distance along this diagonal from the charge sector boundary.

In the vicinity of the transition, the lowest-energy states of the DQD are spanned by a basis $\lbrace \vert L \rangle, \vert R \rangle \rbrace$, states having excess charge on the left or right quantum dot, respectively. 
In this basis, the qubit Hamiltonian is $H = -(\epsilon \sigma_{z} + \Delta \sigma_{x})/2$, where $\epsilon$ is the detuning, $\Delta$ is the tunnel coupling between the dots, and $\sigma_{x}$ ($\sigma_{z}$) is the X (Z) Pauli matrix.  
We denote the density matrix of the qubit at time $t$ as $\rho(t)$. 

Following the diagonal path in ($V_{LP}, V_{RP}$) indicated in Fig.  \ref{fig:DQD_SEM_Stability_Diagram}(b), the charge occupation shifts from the left to right QD.
 At thermal equilibrium, the population of the ground and excited states correspond to their respective Boltzmann weights. The probability for the qubit to be in state $\vert R \rangle$ is $P_{R} = \frac{1}{2} \big[ 1 - \frac{\epsilon}{\hbar \Omega} \tanh\big( \frac{\hbar \Omega}{2 k_{B} T_{e}} \big) \big]$, 
where $\hbar \Omega = \sqrt{\epsilon^{2} + \Delta^{2}}$ is the energy gap and $T_{e}$ is the electron temperature, determined to be $T_{e} \approx \mathrm{300 \ mK}$ by fitting this formula to the observed inter-dot charge transition \cite{DiCarlo2004}.  The derivative of this charge distribution with respect to detuning appears as a peak in the equilibrium differential charge sensing data about $\epsilon = 0$, shown in Fig. \ref{fig:DQD_SEM_Stability_Diagram}(c). 

The ground and excited state occupations coincide with their equilibrium values if the rate of relaxation to equilibrium is fast relative to the timescale of any control variation. 
Deviation from this equilibrium charge occupation occurs as control pulses approach relaxation timescales. 
We probe the relaxation rate and energy dependence by varying the pulse timescale and amplitude. 
More precisely, we measure the time-averaged charge occupation $n = \overline{\langle R \vert \rho(t) \vert R \rangle}$ in the presence of a square wave of fixed peak-to-peak amplitude $\delta \epsilon$ and frequency $f$, superimposed onto a variable DC detuning offset $\overline{\epsilon}$ such that the detuning periodically takes the values $\epsilon_{\pm} = \overline{\epsilon} \pm \delta \epsilon/2$. 
If $\epsilon_{-} < 0 < \epsilon_{+}$, the qubit repeatedly passes through the anti-crossing at $\epsilon=0$, with a waiting time $t_{w}$ between each detuning sweep. 

If the Hamiltonian is swept through $\epsilon=0$ in a sufficiently short time $\tau$, ground state population will be pumped into excited state population and vice versa. 
The transition will be diabatic if $\tau \ll 2 \hbar \delta \epsilon / \pi \Delta^{2}$, according to the Landau-Zener formula \cite{Landau1932, Zener1932, Majorana1932, Stueckelberg1932}.  Since the ramp time $\tau$ is held fixed in this experiment ($\tau \approx 16 \ \mathrm{ns}$) and a change in the observed $n$ is effected solely through variation of $f$, we argue that the transitions in this device are diabatic, i.e. the sweep between $\epsilon_{\pm}$ is fast compared with the adiabatic timescale for traversing the avoided level crossing. 
If the transitions between $\epsilon_{-}$ and $\epsilon_{+}$ were adiabatic, we would expect to observe that $n(\overline{\epsilon}, f)$ is independent of $f$. 

\paragraph{Experimental results-}
\begin{figure}[h]
\includegraphics[width=0.48\textwidth]{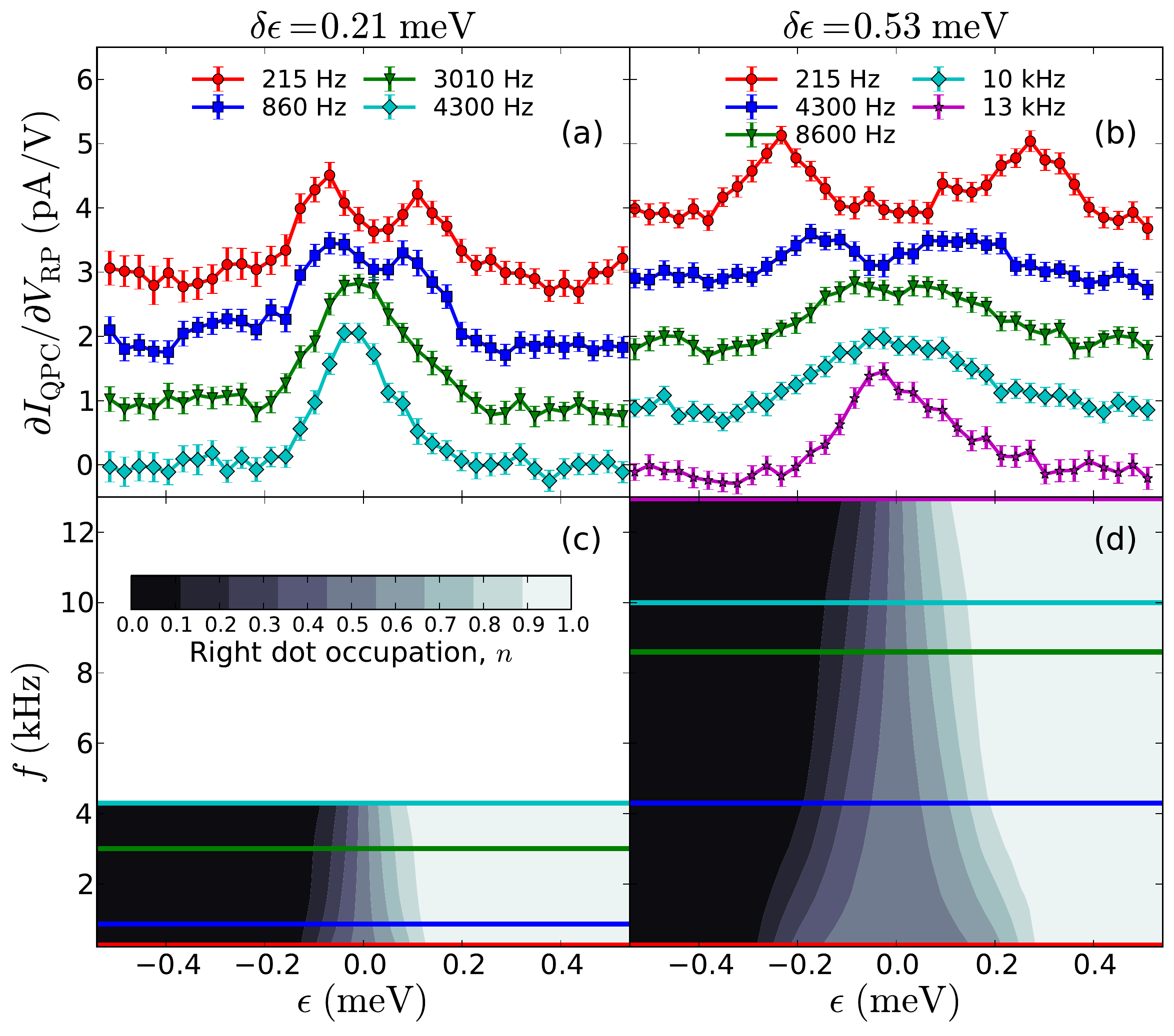}
\vspace{-20pt}
\caption{(Color online) Differential charge sensing measurements for toggling amplitudes of (a) 0.21 meV and (b) 0.53 meV. Error bars correspond to $\pm$ twice the standard error of the mean, with 100 measurements at each detuning value. 
The time-averaged occupation $n(\overline{\epsilon}, f)$ for $\delta \epsilon$ equal to (c) 0.21 meV and (d) 0.53 meV, calculated from smoothed differential conductance data [see Supplementary].  Horizontal lines denote the frequencies probed in the experiment.}
\vspace{-10pt}
\label{fig:Diff_and_occupation_2x2Plot}
\end{figure}

Figures \ref{fig:Diff_and_occupation_2x2Plot}(a,b) show differential charge measurements with this square wave modulation of the detuning. 
For small $f$, Figs. \ref{fig:Diff_and_occupation_2x2Plot}(a,b) exhibit two peaks, corresponding approximately to the two detuning locations for which either $\epsilon_{+}$ or $\epsilon_{-}$ align with the zero detuning point. 
Toggling at this low frequency leads to a doubled charge stability diagram, i.e. two superimposed copies of the equilibrium diagram as in Fig. \ref{fig:DQD_SEM_Stability_Diagram}(c) shifted relative to one another along the diagonal by the toggling amplitude $\delta \epsilon$. 
The separation of the peaks for small $f$ is commensurate with the peak-to-peak amplitude of the square wave. 
When two peaks are well defined in the QPC transconductance, this indicates distinct regions of zero, half, and one average charge occupation. 
In this regime, relaxation is sufficiently fast that all excited state occupation has relaxed to the ground state well before the waiting time $t_{w}$. 
The intermediate plateau, where $n \approx 0.5$, is visible in Fig. \ref{fig:Diff_and_occupation_2x2Plot}(c,d). 
The peaks merge as $f$ increases, indicating a significant divergence from the thermal equilibrium distribution. 
This signals that the waiting time is approaching the relaxation timescale. 
As $f$ grows, the charge qubit does not have time to relax completely to its ground state during each waiting interval $t_{w}$, and as a result the middle plateau region disappears. 
This variation of the mean charge distribution with $f$ and $\overline{\epsilon}$ provides quantitative information about how the relaxation rate depends on the detuning energy.
We observe that the frequency at which the peaks merge depends on the toggling amplitude $\delta \epsilon$, a further indication that the relaxation rate depends on the probe energy scale.  
We furthermore note that the charge sensor is operating in the weak measurement regime where the relaxation times, although slow, are still much faster than the back-action of the charge sensor. 
Therefore, back-action does not significantly perturb the spontaneous emission rates. 
The measurement regime is dependent on the relative rate of relaxation compared to the back-action rate, which has been estimated as $\gamma =(\sqrt{I_{1}}-\sqrt{I_{2}})^2 / 2 \pi e$ \cite{Pioro-Ladriere2005}. 
In this experiment the change in sensor current was of the order of 0.25 pA, amounting to order of Hz back-action compared to kHz relaxation times. 

\begin{figure}[h]
\includegraphics[width=0.48\textwidth]{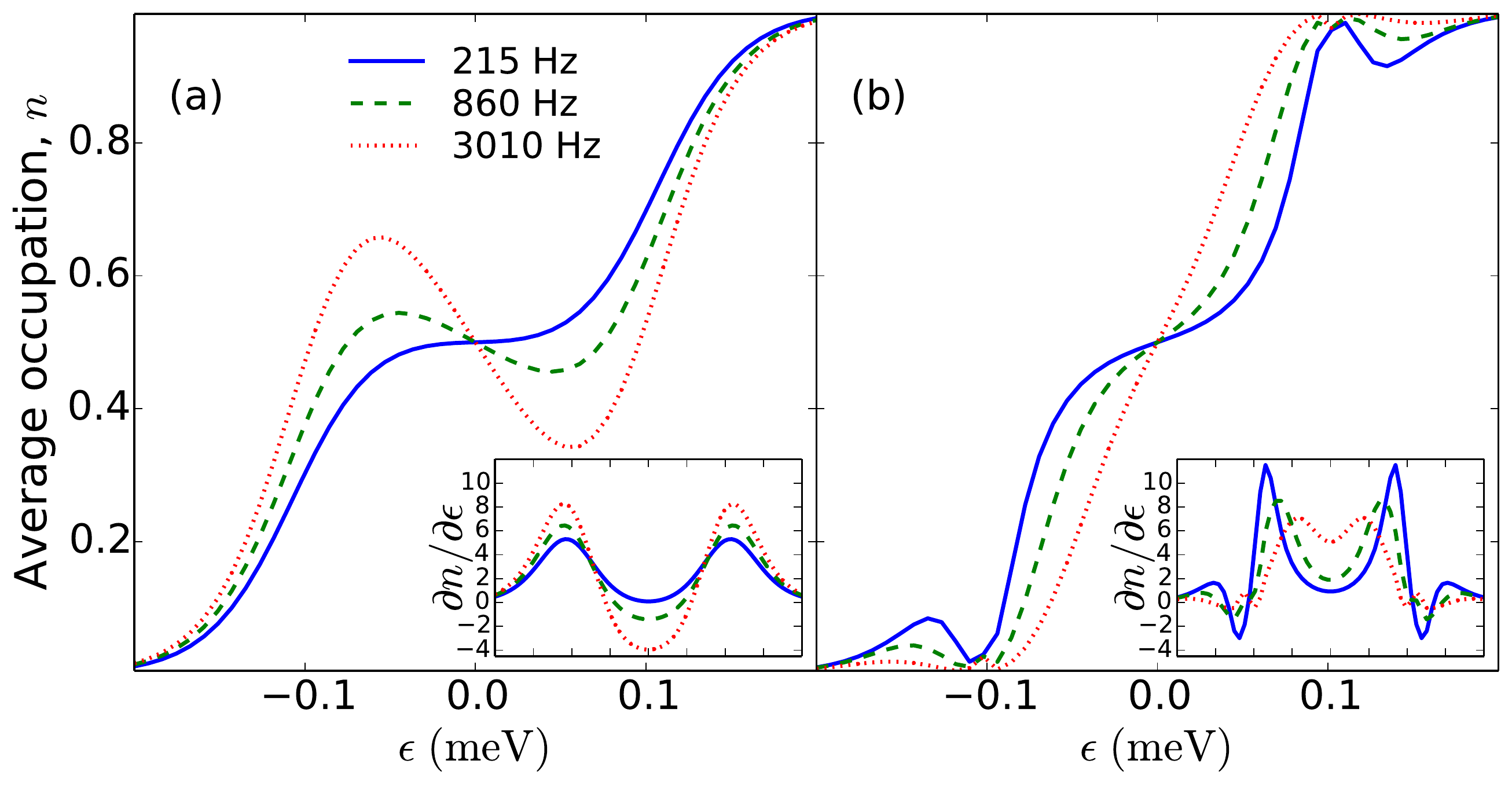}
\vspace{-20pt}
\caption{(Color online) Examples of charge occupation vs detuning according to the phenomenological rate model, taking $\delta \epsilon=0.21 \ \mathrm{meV}$, $\Delta = 1 \ \mathrm{\mu eV}$, $T=0.3\ \mathrm{K}$, and $\hbar \omega_{c}=0.5\ \mathrm{meV}$, for (a) Ohmic spectral density ($s=1$, $\alpha=10^{-4}$) and (b) super-Ohmic spectral density ($s=5$, $\alpha=0.2$).  Insets are the differential of the charge occupation curves.  Merging of the peaks, as in the experiment, is a signature of super-Ohmic behavior, and the shape of the merging is sensitive to the specific energy dependence of the spectral density function.}
\vspace{0pt}
\label{fig:OccupationOhmicVsSuperOhmic}
\end{figure}

\paragraph{Rate modeling-}  
We model the coupling between the charge qubit and its environment according to the spin-boson model \cite{Leggett1987}.  
Making the Born and Markov approximations, where the coupling to the bath is assumed to be sufficiently weak and the bath timescales shorter than any relevant qubit timescales \cite{Breuer2002, Eckel2006}, the rate of relaxation to thermal equilibrium depends on the tunnel coupling $\Delta$, energy gap $\hbar \Omega$, and spectral density $J(\omega)$ of the boson bath as
\be
\Gamma_{r}(\omega) = \frac{2 \pi}{\hbar^{2}} \Big( \frac{\Delta}{\hbar \omega} \Big)^{2} J(\omega) \coth(\beta \hbar \omega / 2),
\label{eq:RelaxationRate}
\ee
where $\beta = (k_{B} T)^{-1}$ is the inverse temperature \cite{Breuer2002} [see Supplementary]. 
The response of the calculated electron occupation depends strongly on the functional form of the spectral density, as illustrated in Fig \ref{fig:OccupationOhmicVsSuperOhmic} by the contrast between two phenomenological example cases, Ohmic and super-Ohmic. 
From Eq. (\ref{eq:RelaxationRate}) we can extract properties of $J(\omega)$ by fitting to the experimentally observed charge occupations as a function of the detuning parameters and pulse frequency $f$. 
The dependence of the relaxation rate on the energy gap $\hbar \omega$ can be deduced quantitatively, as described below, even in regimes where the relaxation time is much longer than either the dephasing time \cite{Hayashi2003} or transport times through the DQD \cite{Fujisawa1998}.

\begin{figure}[h]
\includegraphics[width=0.45\textwidth]{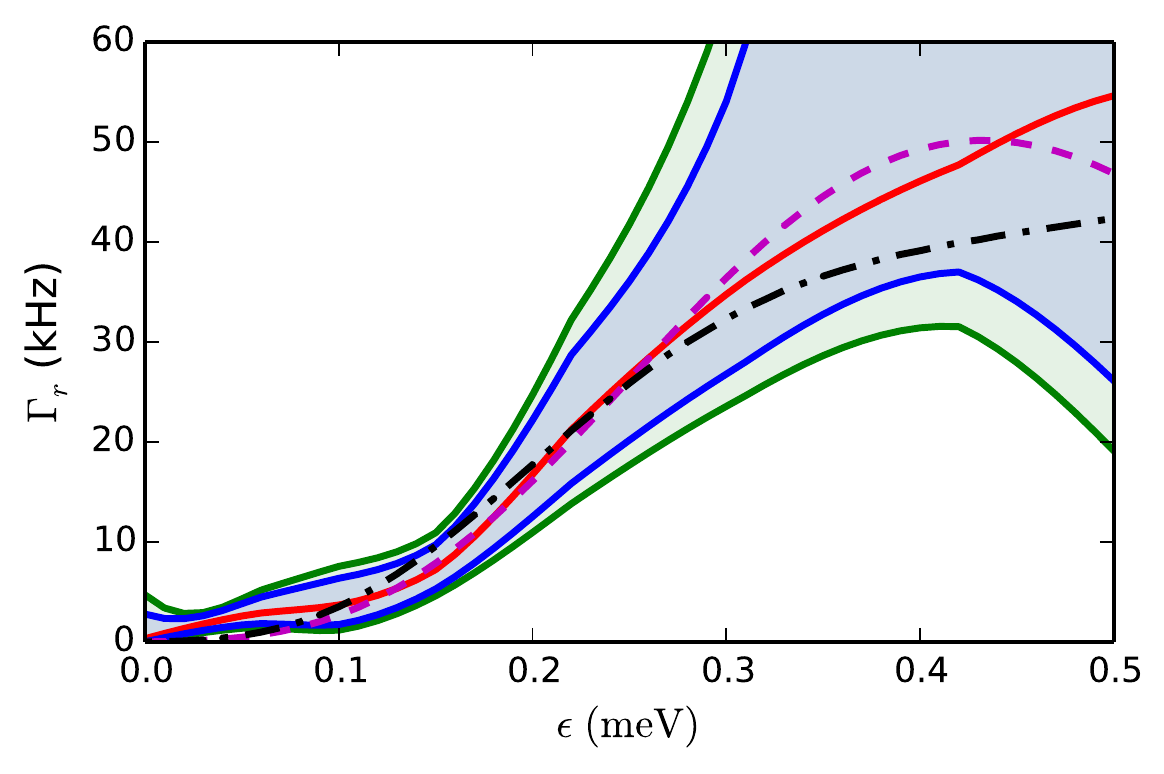}
\vspace{-10pt}
\caption{(Color online) Relaxation rate $\Gamma_{r}(\epsilon)$. The green (blue) pairs of curves bound the shaded 68\% (95\%) confidence region, while the best-fit relaxation rate is shown in red (solid).  The best-fit phenomenological relaxation rate is plotted in magenta (dashed, fit parameters: $s=5$, $\hbar \omega_{c} = 0.25 \ \mathrm{meV}$, and $\alpha = 0.29$, assuming $\Delta = 1 \ \mathrm{\mu eV}$), while the best fit derived from the simple microscopic model is shown in black (dot-dashed, fit parameters: $L=45 \ \mathrm{nm}$ and $E_{0} = 1.7 \ \mathrm{meV}$).}
\label{fig:RelaxationVsDetuning}
\end{figure}

\paragraph{Fit to experiment-}
In order to fit the relaxation rate to the experiment, we perform a gradient minimization over the space of relaxation rate functions $\Gamma_{r}(\epsilon)$ of the mismatch between the calculated and observed $n(\overline{\epsilon},f)$ [see Supplementary]. 
We seed the optimization with a constant relaxation rate  $\Gamma_{r}(\epsilon) = 10 \ \mathrm{kHz}$ to avoid bias towards any specific spectral density. 
Fig. \ref{fig:RelaxationVsDetuning} shows the best fit, with associated confidence regions. 

We can compare the extracted relaxation rates to what is predicted by a phenomenological spectral density. 
The phenomenological form of the spectral density we consider takes the form $J_{\mathrm{ph}}(\omega)=\alpha \hbar^{2} \omega (\omega / \omega_{c})^{s-1} \exp{-\omega^{2}/2\omega_{c}^{2}}$, where  $\omega_{c}$ is a high-frequency cutoff and $\alpha$ is a unitless parameterization of the qubit-bath coupling strength. 
Reasonably good agreement is found for the case $s=5$, consistent with an acoustic phonon dominated mechanism.

\paragraph{Microscopic model}
To test quantitative agreement with an acoustic phonon mechanism, we derive the spectral density function $J(\omega)$ for a simple microscopic model of the DQD and DQD-phonon interaction, taking the basis states for the DQD to be the Fock-Darwin ground states of anisotropic harmonic oscillator potentials localized to the left and right wells. 
A similar microscopic model has been considered in Ref. \cite{Wang2013}.
We parameterize this system in terms of two parameters: the confinement energy $E_{0}$ and the dot-dot separation $2L$.  
This more detailed microscopic model leads to a more complex expression for $J(\omega)$ which depends on the system dimensions and various material parameters [see Supplementary]. 
The best fit dot dimensions of the DQD are a dot diameter $2a \approx 30 \ \mathrm{nm}$ and dot-dot separation $2L \approx 90 \ \mathrm{nm}$. 
This is consistent with the lithographic size of the dot and the measured gate capacitances to the dots. 
We find $\Delta \approx 1 \ \mathrm{\mu eV}$ from fitting the microscopic model to the observed $n(\overline{\epsilon},f)$, with only confinement energy $E_{0}$ and dot-dot separation $2L$ as the fit parameters [see Supplementary].

\paragraph{Summary-}
In summary, we have measured the energy-dependent rate of spontaneous emission in a MOS Si double quantum dot over a wide range of energies.  
We note that this pulsed measurement approach for extracting the spectral density function is generally applicable to other two-level systems, such as flux qubits. 
For the system studied here, slow relaxation rates are discernible because the charge in the DQD is well isolated from the leads and is measured in the weak measurement limit, minimizing the effect of back-action. 
A super-Ohmic spectral density consistent with an acoustic phonon mechanism of inelastic relaxation best fits the experiment.  
Our technique also provides an estimate of the tunnel coupling $\Delta$, which is otherwise difficult to characterize in the weak tunnel coupling regime ($\Delta < k_{B}T$). 
This regime is of particular interest to quantum annealing schemes, for which understanding the noise dynamics during slow evolutions through narrow gaps is essential to predicting the performance of such computing devices.

\paragraph{Methods-}
Experiment:
A global Al top gate induces a two-dimensional electron gas (2DEG) near the Si/Si$\mathrm{O}_{2}$ interface. The lower polysilicon gates deplete the 2DEG, forming a DQD between gates L, TP, CP, and R. 
We bias the QPC with 400 $\mu$V DC and apply a 200 $\mu$V AC modulation at 381 Hz to monitor the differential conductance of the constriction with a standard lock-in technique. 
Modulating the LP and RP gates in phase with a 2 mV AC signal at a frequency of 43 Hz further enhances the charge sensing signal. 
Gates LP and RP are connected to coaxial cables with a measured bandwidth exceeding 10 MHz.  
An Agilent 33522A arbitrary waveform generator produces square wave pulses with a rise time of 16 ns, applied concurrently to the LP and RP gates. 
All measurements presented here are time-averaged over $T_{\mathrm{av}} = 300\ \mathrm{ms}$.  

Model:
We model the periodic controls and resulting dynamics of the excited and ground state populations in this experiment according to a piecewise-defined rate equation. 
The rate of change of population in the energy basis is governed by coupled differential equations describing the detailed balance resulting from Eq. \ref{eq:RelaxationRate} and $\epsilon(t)$ for each detuning value. 
Harbusch, et al. have pursued a similar rate equation approach in Ref. \cite{Harbusch2010b}. 
As a result of the periodicity of the controls being much shorter than the averaging time and the averaging time being much longer than all other relevant timescales in the system, it is reasonable to assume the system to have dynamically equilibrated well before $T_{\mathrm{av}}$. 
With this assumption, we compute the time-averaged charge occupation $n(\bar{\epsilon}, f)$ [see Supplementary]. 

\paragraph{Acknowledgements-} 
We thank Andy Sachrajda, Michel Pioro-Ladri\`ere, Jonathan Moussa, Wayne Witzel, and John King Gamble for fruitful discussions. 
This work was supported by the Laboratory Directed Research and Development program at Sandia National Laboratories and performed, in part, at the Center for Integrated Nanotechnologies, a U.S. Department of Energy Office of Basic Energy Sciences user facility. Sandia is a multiprogram laboratory operated by Sandia Corporation, a Lockheed Martin Company, for the United States Department of Energy under Contract No. DE-AC04-94AL85000.

\bibliography{mybibliography}
\end{document}


\title{Supplementary Information}

\author{Khoi T. Nguyen}
\author{N. Tobias Jacobson}
\author{Michael P. Lilly}
\author{N. Bishop}
\author{E. Nielsen}
\author{J. Wendt}
\author{J. Dominguez}
\author{T. Pluym}
\author{Malcolm S. Carroll}
\email{mscarro@sandia.gov}
\affiliation{Sandia National Laboratories, Albuquerque, New Mexico 87185, USA}

\maketitle

\section{Deriving the relaxation rate}
We model the charge DQD system, its environment, and the interaction between the system and its environment as a spin-boson model \cite{Leggett1987}. 
The system Hamiltonian $H_{S}$, environment (bath) Hamiltonian $H_{B}$, and the system-environment interaction $H_{I}$, respectively, are 
\bea
H_{S} & = & -\frac{1}{2}(\epsilon\sigma_{z}+\Delta\sigma_{x}) \\
H_{B} & = & \sum_{k}\hbar\omega_{k}\hat{b}_{k}^{\dagger}\hat{b}_{k} \\
H_{I} & = & \sigma_{z}\sum_{k}\big(g_{k}\hat{b}_{k}^{\dagger}+g_{k}^{*}\hat{b}_{k}\big),
\eea
where $\hat{b}_{k}^{\dagger}$ ($\hat{b}_{k}$) creates (annihilates) the $k$-th mode of a bosonic degree of freedom with energy $\hbar \omega_{k}$, and $g_{k}$ parameterizes the strength of coupling between the system and the $k$-th bosonic mode. 
The coupling parameters $g_{k}$ depend primarily on the DQD geometry and bulk properties of the semiconductor in which the DQD is embedded. 
As we will argue later, in the system we study here the coupling between the DQD and its environment is mediated primarily through $\sigma_{z}$. 
If the tunnel coupling, $\Delta$, is sufficiently strong, however, it may be necessary to include coupling to the environment through $\sigma_{x}$ as well. 

The ground and first-excited eigenstates of $H_{S}$ are 
\bea
\vert E_{0}\rangle & = & \cos(\theta)\vert L\rangle+\sin(\theta)\vert R\rangle \nonumber \\
\vert E_{1}\rangle & = & \sin(\theta)\vert L\rangle-\cos(\theta)\vert R\rangle,
\eea
where $\theta=\frac{1}{2}\arctan(\Delta/\epsilon)\in[0,\pi/2]$, and the energy gap is $\hbar \Omega = \sqrt{\epsilon^{2} + \Delta^{2}}$. 
In our labeling convention, the ground state tends to the localized basis state $\vert L\rangle$ ($\vert R\rangle$) for $\epsilon / \hbar \Omega \to \infty$ ($\epsilon / \hbar \Omega \to -\infty$).

Making the standard Born and Markov approximations \cite{Breuer2002}, we obtain the following equation of motion (in the interaction picture) for the ground state occupation, $\rho_{00}(t) = \langle E_{0} \vert \rho(t) \vert E_{0} \rangle$,
\be
\dot{\rho}_{00} (t) = \Gamma(\Omega)\vert\langle E_{0}\vert\sigma_{z}\vert E_{1}\rangle\vert^{2}\big[1-(1+e^{-\beta\hbar\Omega})\rho_{00} (t)\big)\big],
\ee
where $\vert\langle E_{0}\vert\sigma_{z}\vert E_{1}\rangle\vert^{2}=\big(\Delta/\hbar\Omega\big)^{2} $ and 
\bea
\Gamma(\omega) & = & \frac{1}{\hbar^{2}} \int_{-\infty}^{\infty} \! \ud s \ \mathrm{Tr} \big[ \rho_{B} e^{i H_{B} s / \hbar} B e^{-i H_{B} s / \hbar} B \big] \nonumber \\
& = & \frac{2\pi}{\hbar^{2}}\big[(1+N(\hbar\omega))J(\omega)+N(-\hbar\omega)J(-\omega)\big],
\eea
where $B = \sum_{k}\big(g_{k}\hat{b}_{k}^{\dagger}+g_{k}^{*}\hat{b}_{k}\big)$ is the bath part of the system-bath interaction and $\rho_{B} = e^{-\beta H_{B}}/\mathrm{Tr}(e^{-\beta H_{B}})$ is the thermal state of the bath at inverse temperature $\beta = 1/k_{B} T$. 
The spectral density function is $J(\omega) = \sum_{k}\vert g_{k}\vert^{2}\delta(\omega-\omega_{k})$, and $N(E)=(e^{\beta E}-1)^{-1}$ is the Bose-Einstein distribution.

Finally, the rate of relaxation towards equilibrium takes the form
\bea
\Gamma_{r}(\Omega) & = & \Gamma(\Omega)\vert\langle E_{0}\vert\sigma_{z}\vert E_{1}\rangle\vert^{2} (1+e^{-\beta\hbar\Omega}) \nonumber \\
& = & \frac{2\pi}{\hbar^{2}}\big(\frac{\Delta}{\hbar\Omega}\big)^{2}J(\Omega)\coth(\frac{\beta}{2}\hbar\Omega).
\eea

\section{Experiment}
We have determined the lever arm, relating the detuning voltage to the effective detuning bias $\epsilon$, to be $\alpha = 0.021 \ \mathrm{eV/V} \pm 10\%$ from bias triangle measurements. 
Due to the diagonal sweep in the $(V_{RP}, V_{LP})$ plane, the detuning voltage is larger than either $V_{LP}$ or $V_{RP}$ by a factor of approximately $\sqrt{2}$.  
Explicitly, the detuning signal is $\epsilon(t) = \overline{\epsilon} + (\delta \epsilon / 2) h(f t) + A \sin(2 \pi \nu t) $, where $A \approx 0.06 \ \mathrm{meV}$ due to the $4 \ \mathrm{mV}$ peak-to-peak (in $V_{RP}$) modulation, $\nu = 43 \ \mathrm{Hz}$, and $h(x)$ is a square wave oscillating between $\pm 1$ with period $1$. 
We take $f$ to be a multiple of 43 Hz, ranging from $215 \ \mathrm{Hz}$ to $13 \ \mathrm{kHz}$, and the ramp time $\tau$ always to be much smaller than the waiting time $t_{w} \approx 1/2f$ so that the control pulse is nearly square. 
In addition to varying the toggling frequency $f$ and detuning offset $\overline{\epsilon}$, we consider $\delta \epsilon \in \lbrace 0.21, 0.53 \rbrace \ \mathrm{meV}$. 

\section{Rate equation}
In order to model the experiment, we're interested in evaluating the charge occupation of the system as averaged over a time $T_{\mathrm{av}} \approx 300 \ \mathrm{ms}$. 
This timescale is much longer than all others in the problem.

We incorporate this modulated detuning into our rate equation model by approximating the modulated toggling as a sequence of many piecewise-constant detuning values of duration $\delta t$. 
We then piece together these short intervals over a full period of the controls $t \in [0, T]$, where in our experiment $T ^{-1}= 43 \ \mathrm{Hz}$. 
In the following, we assume that dephasing time in the local energy basis occurs much more rapidly than the duration of each piecewise-constant interval, $\delta t$. 
We also assume that the tunnel coupling $\Delta$ is fixed. 

First, we find the fixed point of the evolution of the full period. 
Each interval corresponds to a map of the form $\mathbf{B}(\epsilon(t+\delta t), \epsilon(t)) \mathbf{R}(\epsilon(t), \delta t)$, where the $2 \times 2$ matrix $\mathbf{R}$ describes the relaxation over the interval $\delta t$ and the $2 \times 2$ matrix $\mathbf{B}(\epsilon(t+\delta t), \epsilon(t))$ performs a change of basis from the energy eigenbasis of $H_{S}(t)$ to that of $H_{S}(t+\delta t)$. 
More precisely,
\be
\mathbf{B}(\epsilon_{1}, \epsilon_{0}) = 
\left(\begin{array}{cc}
\mu(\epsilon_{1}, \epsilon_{0}) & 1 - \mu(\epsilon_{1}, \epsilon_{0})\\
1 - \mu(\epsilon_{1}, \epsilon_{0}) & \mu(\epsilon_{1}, \epsilon_{0})
\end{array}\right),
\ee
where $\mu(\epsilon_{1}, \epsilon_{0}) = \vert \langle E_{0}(\epsilon_{1}) \vert E_{0}(\epsilon_{0})\rangle \vert^{2}$, and 
\be
\mathbf{R}(\epsilon, \delta t) = 
\left(\begin{array}{cc}
\rho_{00}^{\mathrm{eq}}(\epsilon) + \nu \rho_{11}^{\mathrm{eq}}(\epsilon)  & (1-\nu) \rho_{00}^{\mathrm{eq}}(\epsilon)\\
(1 - \nu) \rho_{11}^{\mathrm{eq}}(\epsilon) & \rho_{11}^{\mathrm{eq}}(\epsilon) + \nu \rho_{00}^{\mathrm{eq}}(\epsilon)
\end{array}\right),
\ee
where $\rho_{00}^{\mathrm{eq}}(\epsilon) = (1+e^{-\beta \hbar \Omega(\epsilon)})^{-1} = 1 - \rho_{11}^{\mathrm{eq}}(\epsilon)$ is the thermal equilibrium ground state occupation, $\Gamma_{r}(\epsilon)$ is the rate of relaxation to equilibrium at detuning $\epsilon$, and $\nu = e^{-\delta t \Gamma_{r}(\epsilon)}$. 
Again, we assume that the coherence $\rho_{01} = \langle E_{0} \vert \rho \vert E_{1}\rangle$ in the energy basis vanishes rapidly as compared with the time interval $\delta t$. 

Concatenating all of these intervals together, we obtain the full map 
\be
\mathbf{\Lambda}_{N}= \overleftarrow{\prod}_{k=1}^{N} \mathbf{B}(\epsilon(t_{k+1}), \epsilon(t_{k})) \mathbf{R}(\epsilon(t_{k}), \delta t),
\ee
where $N = T / \delta t$, $t_{k} = (k-1/2) \delta t$, $\mathbf{B}(\epsilon(t_{N+1}),\epsilon(t_{N})) \equiv \mathbf{B}(\epsilon(t_{1}),\epsilon(t_{N}))$ due to control periodicity, and the product is time-ordered from right to left. 
Let the state $\rho(0)$ be the fixed point of the map $\mathbf{\Lambda}_{N}$, i.e. $\rho(0) = \mathbf{\Lambda}_{N} \rho(0)$.
We assume that the time average is evaluated over a sufficiently long timescale $T_{\mathrm{av}}$ that all information about the initial conditions is washed out and the system reaches a dynamical equilibrium on a timescale much shorter than $T_{\mathrm{av}}$.
To evaluate the full time average, then, we can take $\rho(0)$ as the initial state and evaluate the average charge occupation over the interval $t \in [0,T]$. 
We approximate the switching between the constant-detuning intervals as instantaneous. 

Given a ground state occupation $\rho_{00}$, the corresponding expectation for the charge to be in the left well is
\be
n_{L}(\epsilon) = \frac{1}{2}\Big[ 1 - \frac{\epsilon}{\sqrt{\epsilon^{2} + \Delta^{2}}} \Big] + \frac{\epsilon}{\sqrt{\epsilon^{2} + \Delta^{2}}} \rho_{00}(\epsilon).
\ee

The time-averaged occupation of the left well is
\bea
\overline{n_{L}} & = & \frac{1}{T} \int_{0}^{T} d t' n_{L}(\epsilon(t')) \nonumber \\
& = & \frac{1}{T} \sum_{k=1}^{N} \int_{(k-1) \delta t}^{k \delta t} d t' n_{L}(\epsilon(t')) \label{eq:MeanOcc}.
\eea
For $(k-1) \delta t < t < k \delta t$, 
\bea
\rho(t) & = & \mathbf{R}(\epsilon(t_{k}), t \! - \! (k \! - \! 1)\delta t) \rho((k \! - \! 1) \delta t) \nonumber \\
& = & \mathbf{R}(\epsilon(t_{k}), t \! - \! (k \! - \! 1)\delta t) \mathbf{\Lambda}_{k-1} \rho(0),
\eea
hence the average state over this same $k^{\mathrm{th}}$ interval is
\be
\overline{\rho}^{(k)} = \overline{\mathbf{R}}(\epsilon(t_{k}), \delta t) \mathbf{\Lambda}_{k-1} \rho(0),
\ee
where $\mathbf{\Lambda}_{0} \equiv \1$ and
\be
\overline{\mathbf{R}}(\epsilon, \delta t) = 
\left(\begin{array}{cc}
\rho_{00}^{\mathrm{eq}}(\epsilon) + \overline{\nu} \rho_{11}^{\mathrm{eq}}(\epsilon)  & (1 - \overline{\nu}) \rho_{00}^{\mathrm{eq}}(\epsilon)\\
(1 - \overline{\nu}) \rho_{11}^{\mathrm{eq}}(\epsilon) & \rho_{11}^{\mathrm{eq}}(\epsilon) + \overline{\nu} \rho_{00}^{\mathrm{eq}}(\epsilon)
\end{array}\right),
\ee
with $\overline{\nu} = (1 - e^{-\delta t \Gamma_{r}(\epsilon)})/(\delta t \Gamma_{r}(\epsilon))$.
From this, Eq. (\ref{eq:MeanOcc}), and the piecewise-constant approximation we can compute $\overline{n_{L}}$ as
\be
\overline{n_{L}} = \frac{1}{N} \sum_{k=1}^{N} \frac{1}{2} \Big[ 1 - \frac{\epsilon_{k}}{\sqrt{\epsilon_{k}^{2} + \Delta^{2}}} \Big] + \frac{\epsilon_{k}}{\sqrt{\epsilon_{k}^{2} + \Delta^{2}}} \overline{\rho}^{(k)}_{00},
\ee
where $\epsilon_{k} = \epsilon(t_{k})$.

\section{Model for DQD as a harmonic double well potential}
We model the double wells of the DQD as a pair of identical harmonic potentials, given by \cite{Nielsen2010}
\be
V(\mathbf{r}, \epsilon) = \min \big\lbrace \alpha_{x} (x + L)^{2}, \alpha_{x} (x - L)^{2} + \epsilon  \big\rbrace + \alpha_{y} y^2 + \alpha_{z} z^2,
\ee
where $\alpha_{x} = \alpha_{y} = m_{\perp} E_{0}^{2} / 2 \hbar^{2}$, $\alpha_{z} = m_{\parallel} E_{z}^{2} / 2 \hbar^{2}$. 
The confinement energies $E_{0}$ and $E_{z}$ give a characteristic dot width $a = \sqrt{\hbar^{2} / m_{\perp} E_{0}}$ and thickness $b = \sqrt{\hbar^{2} / m_{\parallel} E_{z}}$. 
We make the single-valley approximation, which should be valid for sufficiently strong confinement to the interface perpendicular to the $z$-axis. 
That is, we assume confinement in the $z$-direction such that $b \ll a$. 
Here, $m_{\perp}$ ($m_{\parallel}$) is the effective mass corresponding to the axes perpendicular (parallel) to the principal axis of the given valley. For silicon, $m_{\perp}=0.19m_{e}$ and $m_{\parallel} = 0.98 m_{e}$ \cite{RidleyBook}.

In addition to this pair of harmonic potentials, as in the experiment let there also be a magnetic field $\mathbf{B}$ oriented perpendicularly to the interface, $\mathbf{B}=B\hat{z}$. 
The eigenstates of a single harmonic potential in a magnetic field are given by the Fock-Darwin states \cite{Wang2011}.
In particular, we're interested in the ground state, as we will assume the left/right basis to be given by the ground states of the respective quantum dots. 
Following Ref. \cite{Wang2011}, the ground Fock-Darwin states for the left and right wells are given by
\bea
\varphi_{L/R}(\mathbf{r}) & = & \frac{1}{\sqrt{\pi}l_{0}}\exp\big[\pm i \Big(\frac{eBL}{2\hbar} \Big)y\big]\exp\big[-\frac{(x \pm L)^{2}+y^{2}}{2l_{0}^{2}}\big] \nonumber \\
& \times & \frac{1}{\sqrt{\pi^{1/2}b}}\exp\big[-\frac{z^{2}}{2b^{2}}\big], \label{eq:PhiL}
\eea
where 
\be
l_{0} = \frac{\hbar}{\sqrt{m_{\perp}}}\Big(\big(\frac{\hbar\omega_{c}}{2}\big)^{2}+E_{0}^{2}\Big)^{-1/4}
\ee
and $\omega_{c}=eB/m_{\perp}$ is the Larmor frequency. 
Note that the overlap is: 
\be
s\equiv\langle\varphi_{L}\vert\varphi_{R}\rangle=\exp\big[-\Big(\Big(\frac{L}{l_{0}}\Big)^{2}+\Big(\frac{eBLl_{0}}{2\hbar}\Big)^{2}\big)\big],
\ee
so the magnetic field leads to enhanced confinement and consequently some amount of suppression of the overlap. 
The confinement energy $E_{0}$ determines the dot size, and if the inter-dot separation $2L$ is known the tunnel coupling $\Delta$ can be computed, as detailed in the following section. 

\subsection{Computing the tunnel coupling, $\Delta$}
Given the parameters $E_{0}$, $L$, and $\epsilon$ of the above double well potential, we now describe how the tunnel coupling may be computed. 
Recall that the system Hamiltonian is $H = -(1/2) \big( \epsilon \sigma_{z} + \Delta \sigma_{x} \big)$ in the basis $\lbrace \vert L \rangle, \ \vert R \rangle \rbrace$. 
Hence, the tunnel coupling is $\Delta = -2 \langle L \vert H \vert R \rangle$. 
To evaluate this matrix element, it's convenient to split the Hamiltonian into 
\be
H = \frac{1}{2} \Big( H_{L} + H_{R} \Big) + \delta V,
\ee
where
\bea
H_{L} & = & K + \alpha_{x} (x+L)^{2} + \alpha_{y} y^{2} + \alpha_{z} z^{2} \nonumber \\
H_{R} & = & K + \alpha_{x} (x-L)^{2} + \alpha_{y} y^{2} + \alpha_{z} z^{2} + \epsilon \nonumber
\eea
and
\bea
\delta V & = & \min\Big\lbrace \alpha_{x} (x+L)^{2}, \alpha_{x} (x-L)^{2}+ \epsilon \Big\rbrace \nonumber \\
& - & \frac{1}{2} \Big(\alpha_{x}(x+L)^{2}+\alpha_{x} (x-L)^{2} + \epsilon\Big) \nonumber \\
& = & \mathrm{sgn}(x - x_{0}) \Big( \frac{\epsilon}{2} - 2 \alpha_{x} L x \Big),
\eea
where $x_{0} = \epsilon / 4 \alpha_{x} L$. 
We consider the detuning to be sufficiently small that $-L < x_{0} < L$, so that the harmonic local minima at $x = \pm L$ are well-defined. 
The kinetic term, $K$, is given by
\bea
K & = & \frac{-\hbar^{2}}{2 m_{\perp}} \Big( \frac{\partial^{2}}{\partial x^{2}} + \frac{\partial^{2}}{\partial y^{2}} + \gamma \frac{\partial^{2}}{\partial z^{2}} \Big) \nonumber \\
& + & \frac{i\hbar eB}{4m_{\perp}}\Big(y\partial_{x}-x\partial_{y}\Big)+\frac{e^{2}B^{2}}{8m_{\perp}}(x^{2}+y^{2}),
\eea
where $\gamma = m_{\perp} / m_{\parallel}$. 
Note that the Fock-Darwin states $\varphi_{L/R}(\mathbf{r})$ are the ground states of $H_{L}$ and $H_{R}$, respectively. 
Denote $E_{L} = \langle \varphi_{L} \vert H_{L} \vert \varphi_{L} \rangle$ and $E_{R} = \langle \varphi_{R} \vert H_{R} \vert \varphi_{R} \rangle$. 
Using Eq. (\ref{eq:OrthoBasis}), we may now evaluate the matrix element $\langle L \vert H \vert R \rangle$ in terms of matrix elements of $H$ with respect to the non-orthogonal Fock-Darwin states. 
Noting that 
\be
H_{L} = H_{R} + 4 \alpha_{x} L x - \epsilon,
\ee
we find 
\bea
\langle \varphi_{L} \vert H \vert \varphi_{L} \rangle & = & E_{L} + 2 \alpha_{x} L^{2} + \frac{\epsilon}{2} + \langle \varphi_{L} \vert \delta V \vert \varphi_{L}\rangle \nonumber \\
\langle \varphi_{R} \vert H \vert \varphi_{R} \rangle & = & E_{R} + 2 \alpha_{x} L^{2} - \frac{\epsilon}{2} + \langle \varphi_{R} \vert \delta V \vert \varphi_{R} \rangle \nonumber \\
\langle \varphi_{L} \vert H \vert \varphi_{R} \rangle & = & \frac{1}{2}(E_{R}+E_{L})s + \langle \varphi_{L} \vert \delta V \vert \varphi_{R} \rangle. 
\eea

We now need to evaluate the matrix elements $\langle \varphi_{L} \vert \delta V \vert \varphi_{L}\rangle$, $\langle \varphi_{R} \vert \delta V \vert \varphi_{R} \rangle$, and $\langle \varphi_{L} \vert \delta V \vert \varphi_{R} \rangle$. 
For compactness, we denote these $\delta V_{LL}$, $\delta V_{RR}$, and $\delta V_{LR}$, respectively.
By a straightforward integration, we find 
\bea
\delta V_{LL} & = & -\Big[ \frac{\epsilon}{2} + 2 \alpha_{x} L^{2} \Big] \mathrm{Erf}\Big( \frac{L+x_{0}}{l_{0}} \Big) - \frac{2 \alpha_{x} L l_{0}}{\sqrt{\pi}} e^{-(\frac{L+x_{0}}{l_{0}})^{2}}\nonumber \\
\delta V_{RR} & = & \Big[ \frac{\epsilon}{2} - 2 \alpha_{x} L^{2} \Big] \mathrm{Erf}\Big( \frac{L-x_{0}}{l_{0}} \Big) - \frac{2 \alpha_{x} L l_{0}}{\sqrt{\pi}} e^{-(\frac{L-x_{0}}{l_{0}})^{2}} \nonumber \\
\delta V_{LR} & = & -s \Big[ \frac{\epsilon}{2} \mathrm{Erf}\Big(\frac{x_{0}}{l_{0}}\Big) + \frac{2 \alpha_{x} L l_{0}}{\sqrt{\pi}} e^{-(x_{0}/l_{0})^{2}}  \Big],
\eea
where $\mathrm{Erf}(x) = \frac{2}{\sqrt{\pi}} \int_{0}^{x} e^{-t^{2}} \ud t$ is the error function. 
Finally, using Eq. (\ref{eq:OrthoBasis}), the tunnel coupling is
\bea
\Delta & = & \frac{-2}{1 - 2 s g + g^{2}} \Big[ \frac{1}{2} (E_{L} + E_{R})(s(1+g^{2}) - 2 g) \nonumber \\ 
& + & (1+g^{2}) \delta V_{LR} - g \big(4 \alpha_{x} L^{2} + \delta V_{LL} + \delta V_{RR} \big) \Big].
\eea

For $B=100\ \mathrm{mT}$, as in the experiment, the Larmor energy is $\hbar eB/m_{\perp} \approx 60 \ \mathrm{\mu eV}$, which is negligible compared to the confinement energy $E_{0}$ of order meV consistent with the experiment. 
Hence, it is justified to neglect the influence of the magnetic field, and we take $l_{0} \approx a$.

Note that the expression for $\Delta$ includes a dependence on the detuning energy, $\epsilon$. 
We find that for the range of confinement energies $E_{0}$ and dot-dot separations $2L$ that fit this experiment the variation of $\Delta$ with $\epsilon$ is small, less than 1 \% over the range of detunings probed. 
This supports the simplifying assumption of a constant tunnel coupling. 
Given the confinement energy $E_{0}$ and dot-dot separation $2L$, we can now use the results of the next section to compute the spectral density function, $J(\omega)$, and the rate of relaxation to thermal equilibrium, $\Gamma_{r}(\epsilon)$.

\section{A microscopic model for the spectral density}
Following Ref. \cite{RidleyBook}, the Hamiltonian describing the interaction between electrons in the conduction band and phonons is given by
\be
H_{\mathrm{ep}}=\sum_{ij}\Xi_{ij}S_{ij},
\ee
where $\Xi_{ij}$ is a deformation potential tensor and $S_{ij}$ a strain tensor,
\be
S_{ij}=S_{ji}=\frac{1}{2}\Big(\frac{\partial u_{i}}{\partial x_{j}}+\frac{\partial u_{j}}{\partial x_{i}}\Big),
\ee
where $\mathbf{R}=(x_{1},x_{2},x_{3})$ is the position vector of the unit cell and $\mathbf{u}$ is the displacement. 
Expanding the components of the strain tensor in terms of plane waves \cite{RidleyBook}, 
\be
S_{ij}=\frac{1}{2\sqrt{N}}\sum_{\mathbf{q}}\Big[iQ_{\mathbf{q}}(e_{i}q_{j}+e_{j}q_{i}) e^{i\mathbf{q}\cdot\mathbf{R}} +\mathrm{c.c.}\Big],
\ee
where $N$ is the number of unit cells, $\mathbf{e}$ is a unit-length phonon polarization vector, $\mathbf{q}$ is the phonon wavevector, and $\mathrm{c.c.}$ denotes the complex conjugate.  
The normal coordinates $Q_{\mathbf{q}}$ can then be expressed in second-quantized notation in terms of bosonic modes. 
The acoustic branch of these normal modes will have three parts: one longitudinal and two transverse.

The effective deformation potentials for longitudinal modes, $\Xi_{L}(\theta)$, and transverse modes, $\Xi_{T}(\theta)$, are \cite{RidleyBook} 
\bea
\Xi_{L}(\theta) & = & \Xi_{d}+\Xi_{u}\cos^{2}(\theta) \nonumber \\
\Xi_{T}(\theta) & = & \Xi_{u}\sin(\theta) \cos(\theta),
\eea
where $\theta$ is the angle between the phonon wavevector $\mathbf{q}$ and the principal axis of the given valley. 
The $\Xi_{T}(\theta)$ expression is derived in Ref. \cite{RidleyBook} by taking the elastic anisotropy to be small, performing an implicit average over the azimuthal angle of $\mathbf{q}$, and summing over the two transverse acoustic branches. 
The constants $\Xi_{d}$ and $\Xi_{u}$ denote the dilational and uniaxial shear deformation potentials, respectively. 
A remarkable lack of consensus exists in the literature concerning the values of the deformation potentials $\Xi_{d}$ and $\Xi_{u}$ of silicon. 
There is not even agreement as to the \emph{sign} of $\Xi_{d}$ \cite{Fischetti1996, Tahan2002}. 
For example, values of $\Xi_{u}$ have been reported in the range of 7.3 to 10.5 $\mathrm{eV}$, and values of $\Xi_{d}$ in the range -10.7 to 1.1 $\mathrm{eV}$ \cite{Fischetti1996}. 
In this work, we use the values chosen in Ref. \cite{Tahan2002}, $\Xi_{d} = -10.7 \ \mathrm{eV}$ and $\Xi_{u} = 9.29 \ \mathrm{eV}$. 
We note that if the uncertainties and approximations in the following analysis were reduced, our method for determining the relaxation rate may potentially provide an alternative (though indirect) method for estimating the magnitudes of the deformation potentials.

Writing the electron-phonon interaction $H_{\mathrm{ep}} = H_{\mathrm{ep}}^{L} + H_{\mathrm{ep}}^{T}$ in second-quantized notation (i.e. promoting the normal coordinates into bosonic operators), we obtain
\bea
H_{\mathrm{ep}}^{L} & = & \sum_{\mathbf{q}} \xi^{L}(\mathbf{q}) \Big[ e^{i\mathbf{q}\cdot\mathbf{r}} \hat{b}_{L, \mathbf{q}}^{\dagger} - e^{-i\mathbf{q}\cdot\mathbf{r}} \hat{b}_{L, \mathbf{q}} \Big] \nonumber \\
H_{\mathrm{ep}}^{T} & = & \sum_{\mathbf{q}} \xi^{T}(\mathbf{q}) \Big[ e^{i\mathbf{q}\cdot\mathbf{r}} \hat{b}_{T, \mathbf{q}}^{\dagger} - e^{-i\mathbf{q}\cdot\mathbf{r}} \hat{b}_{T, \mathbf{q}} \Big],
\eea
where $\hat{b}_{L, \mathbf{q}}^{\dagger},\hat{b}_{L, \mathbf{q}}$ ($\hat{b}_{T, \mathbf{q}}^{\dagger},\hat{b}_{T, \mathbf{q}}$)
represent the bosonic creation/annihilation operators for longitudinal
(transverse) acoustic phonons and \cite{RidleyBook, MahanBook}
\bea
\xi^{L}(\mathbf{q}) & = & i\sqrt{\frac{\hbar q}{2\rho V c_{L}}}\Big[\big(\Xi_{d}+\Xi_{u}\cos^{2}(\theta)\big)\Big] \nonumber \\
\xi^{T}(\mathbf{q}) & = & i\sqrt{\frac{\hbar q}{2\rho V c_{T}}}\Big[ \Xi_{u}\sin(\theta) \cos(\theta) \Big] \label{eq:xiLongTrans},
\eea
where $\rho = 2.33 \times 10^{3} \ \mathrm{kg/m^{3}}$ is the bulk mass density of silicon and $c_{L} = 9.0 \times 10^{3} \ \mathrm{m/s}$ ($c_{T} = 5.41 \times 10^{3} \ \mathrm{m/s}$) is the speed of sound for longitudinal (transverse) acoustic phonons \cite{Fischetti1996}. $V$ is the unit cell volume, which will cancel out later on in the calculation of the spectral density. 

The electron-phonon interaction takes the form
\begin{eqnarray}
H_{\mathrm{ep}} & = & \sum_{\mathbf{q}, \mu \in \lbrace L, T \rbrace }\xi^{\mu}(\mathbf{q}) e^{i\mathbf{q}\cdot\mathbf{r}} \hat{b}_{\mu, \mathbf{q}}^{\dagger} + \xi^{\mu *}(q) e^{-i \mathbf{q} \cdot \mathbf{r} } \hat{b}_{\mu, \mathbf{q}} \nonumber \\
 & = & \sum_{\mathbf{q},  \mu \in \lbrace L, T \rbrace}\xi^{\mu}(\mathbf{q}) \Big[ \vert L\rangle\!\langle L\vert e^{i\mathbf{q}\cdot\mathbf{r}}\vert L\rangle\!\langle L\vert + \vert L\rangle\!\langle L\vert e^{i\mathbf{q}\cdot\mathbf{r}}\vert R\rangle\!\langle R\vert \nonumber \\
 & + & \vert R\rangle\!\langle R\vert e^{i\mathbf{q}\cdot\mathbf{r}}\vert L\rangle\!\langle L\vert+\vert R\rangle\!\langle R\vert e^{i\mathbf{q}\cdot\mathbf{r}}\vert R\rangle\!\langle R\vert \Big] \hat{b}_{\mu, \mathbf{q}}^{\dagger}+\mathrm{h.c.} \nonumber \\
 & = & \sum_{\mathbf{q},  \mu \in \lbrace L, T \rbrace}\xi^{\mu}(\mathbf{q}) \Big[ \rho_{\mathbf{q}}^{LL}\vert L\rangle\!\langle L\vert+\rho_{\mathbf{q}}^{LR}\vert L\rangle\!\langle R\vert \nonumber \\ 
 & + & \rho_{\mathbf{q}}^{RL}\vert R\rangle\!\langle L\vert+\rho_{\mathbf{q}}^{RR}\vert R\rangle\!\langle R\vert \Big] \hat{b}_{\mu, \mathbf{q}}^{\dagger}+\mathrm{h.c.} \label{eq:HElectronPhonon},
\end{eqnarray}
with
\begin{eqnarray*}
\rho_{\mathbf{q}}^{LL} & = & \langle L\vert e^{i\mathbf{q}\cdot\mathbf{r}}\vert L\rangle \\
\rho_{\mathbf{q}}^{RR} & = & \langle R\vert e^{i\mathbf{q}\cdot\mathbf{r}}\vert R\rangle \\
\rho_{\mathbf{q}}^{LR} & = & \langle L\vert e^{i\mathbf{q}\cdot\mathbf{r}}\vert R\rangle \\
\rho_{\mathbf{q}}^{RL} & = & \langle R\vert e^{i\mathbf{q}\cdot\mathbf{r}}\vert L\rangle,
\end{eqnarray*}
where $\{\vert L\rangle,\vert R\rangle\}$ is an orthonormal basis and $\mathrm{h.c.}$ denotes the Hermitian conjugate. 
To obtain Eq. (\ref{eq:HElectronPhonon}), we have inserted a resolution of the identity, $\1=\vert L\rangle\!\langle L\vert+\vert R\rangle\!\langle R\vert$, on both sides of the exponential factor $e^{i\mathbf{q}\cdot\mathbf{r}}$.
Note that $\rho_{\mathbf{q}}^{RL} = \rho_{\mathbf{q}}^{LR}$ if the wavefunctions are real, which follows from neglecting the magnetic field $\mathbf{B}$.

We can proceed directly to representing this interaction Hamiltonian in the spin-boson representation:
\be
H_{\mathrm{ep}} = \sum_{\mathbf{q}, \mu \in \lbrace L, T \rbrace}(\gamma_{\mathbf{q},\mu}^{z}\sigma_{z}+\gamma_{\mathbf{q},\mu}^{x}\sigma_{x}) \hat{b}^{\dagger}_{\mu, \mathbf{q}} + \mathrm{h.c.},
\ee
where 
\bea
\gamma_{\mathbf{q}, \mu}^{z} & = & \frac{1}{2}\xi^{\mu}(\mathbf{q})\big(\rho_{\mathbf{q}}^{LL}-\rho_{\mathbf{q}}^{RR}\big) \nonumber\\
\gamma_{\mathbf{q}, \mu}^{x} & = & \xi^{\mu}(\mathbf{q})\rho_{\mathbf{q}}^{LR}.
\eea

In the following, we assume that the interfacial confinement to the plane perpendicular to the $z$-axis is sufficiently strong that the single-valley approximation is warranted.
Denoting the (non-orthogonal) pair of Fock-Darwin wavefunctions as
\be
\varphi_{L/R}(\mathbf{r}) = \frac{1}{\sqrt{\pi^{3/2} a^{2} b}}\exp\Big[-\frac{(x\pm L)^{2}+y^{2}}{2a^{2}} \Big] \exp \Big[ -\frac{z^{2}}{2b^{2}} \Big],
\ee
an orthogonal basis $\lbrace \vert L \rangle, \vert R \rangle \rbrace$ in terms of these states is
\bea
\vert L \rangle & = & \frac{1}{\sqrt{1 - 2sg + g^{2}}}\Big( \vert \varphi_{L} \rangle - g \vert \varphi_{R} \rangle \Big) \nonumber \\
\vert R \rangle & = & \frac{1}{\sqrt{1 - 2sg + g^{2}}}\Big( \vert \varphi_{R} \rangle - g \vert \varphi_{L} \rangle \Big) \label{eq:OrthoBasis},
\eea
where $g = (1 - \sqrt{1 - s^{2}})/s$. 
It's convenient to define
\bea
\lambda_{\mathbf{q}}^{LL} & = & \langle \varphi_{L}\vert e^{i\mathbf{q}\cdot\mathbf{r}}\vert \varphi_{L}\rangle \nonumber \\
\lambda_{\mathbf{q}}^{RR} & = & \langle \varphi_{R}\vert e^{i\mathbf{q}\cdot\mathbf{r}}\vert \varphi_{R}\rangle \nonumber \\
\lambda_{\mathbf{q}}^{RL}=\lambda_{\mathbf{q}}^{LR} & = & \langle \varphi_{L}\vert e^{i\mathbf{q}\cdot\mathbf{r}}\vert \varphi_{R}\rangle,
\eea
where
\bea
\lambda_{\mathbf{q}}^{LL} & = & e^{-iLq_{x}}\exp\Big[-\frac{a^{2}(q_{x}^{2}+q_{y}^{2})}{4}\Big]\exp\Big[-\frac{b^{2}q_{z}^{2}}{4}\Big] \nonumber \\
\lambda_{\mathbf{q}}^{LR} & = & e^{-(L/a)^{2}} \exp\Big[-\frac{a^{2}(q_{x}^{2}+q_{y}^{2})}{4}\Big]\exp \Big[-\frac{b^{2}q_{z}^{2}}{4} \Big] \nonumber \\
\lambda_{\mathbf{q}}^{RR} & = & \big(\lambda_{\mathbf{q}}^{LL}\big)^{*}.
\eea
Given Eq. (\ref{eq:OrthoBasis}), 
\bea
\rho_{\mathbf{q}}^{LL} & = & \frac{1}{1 - 2sg + g^{2}} \Big( \lambda_{\mathbf{q}}^{LL} - 2g \lambda_{\mathbf{q}}^{LR} + g^{2} \lambda_{\mathbf{q}}^{RR} \Big) \nonumber \\
\rho_{\mathbf{q}}^{RR} & = & \frac{1}{1 - 2sg + g^{2}} \Big( \lambda_{\mathbf{q}}^{RR} - 2g \lambda_{\mathbf{q}}^{LR} + g^{2} \lambda_{\mathbf{q}}^{LL} \Big) \nonumber \\
\rho_{\mathbf{q}}^{LR} & = & \frac{1}{1 - 2sg + g^{2}} \Big( (1+g^{2}) \lambda_{\mathbf{q}}^{LR} - g( \lambda_{\mathbf{q}}^{LL} + \lambda_{\mathbf{q}}^{RR}) \Big) \nonumber.
\eea

Noting that $g = s/2 +\mathcal{O}(s^{3})$ to lowest order in the overlap $s$, we have 
\bea
\Big\vert \frac{ \rho_{\mathbf{q}}^{LR}}{\rho_{\mathbf{q}}^{LL} - \rho_{\mathbf{q}}^{RR}} \Big\vert & \approx & \Big\vert \frac{\lambda_{\mathbf{q}}^{LR}}{\lambda_{\mathbf{q}}^{LL} - \lambda_{\mathbf{q}}^{RR}} \Big\vert \nonumber \\
& = & \Big\vert \frac{s}{\sin(Lq_{x})} \Big\vert.
\eea
For $L q_{x}$ not close to a multiple of $\pi$ and with the overlap between dot-localized wavefunctions $\vert \varphi_{L/R} \rangle$ small, we should expect that the coupling of the bath through $\sigma_{z}$ will dominate the coupling through $\sigma_{x}$. 
In the following, we keep only the term proportional to $\sigma_{z}$. 

Hence, 
\[
\gamma_{\mu, \mathbf{q}}^{z} \approx -i \xi^{\mu}(\mathbf{q}) \sin(L q_{x}) \exp\Big[\frac{-\big( a^{2}(q_{x}^{2}+q_{y}^{2}) + b^{2}q_{z}^{2} \big)}{4} \Big].
\]
From this and Eq. (\ref{eq:xiLongTrans}), we have
\bea
\vert\gamma_{L, \mathbf{q}}^{z}\vert^{2} & = & \frac{\hbar\vert\mathbf{q}\vert}{2\rho Vc_{L}}\Big[\Xi_{d}+\Xi_{u}\cos^{2}(\theta)\Big]^{2}\sin^{2}(Lq_{x}) \nonumber \\ 
& \times & \exp\Big[-\frac{a^{2}}{2}(q_{x}^{2}+q_{y}^{2})\Big]\exp\Big[-\frac{b^{2}q_{z}^{2}}{2}\Big]
\eea
and
\bea
\vert\gamma_{T, \mathbf{q}}^{z}\vert^{2} & = & \frac{ \hbar \vert \mathbf{q} \vert }{ 2\rho Vc_{T} } \Xi_{u}^{2} \sin^{2}(\theta) \cos^{2}(\theta)\sin^{2}(Lq_{x}) \nonumber \\ 
 & \times & \exp\Big[-\frac{a^{2}}{2}(q_{x}^{2}+q_{y}^{2})\Big]\exp\Big[-\frac{b^{2}q_{z}^{2}}{2}\Big].
\eea
For notational simplicity, we denote $\gamma_{\mu, \mathbf{q}}=\gamma_{\mu, \mathbf{q}}^{z}$. 
To determine the spectral density function, we need to evaluate
\bea
J(\omega) & = & \sum_{\mathbf{q}, \mu \in \lbrace L, T \rbrace}\vert\gamma_{\mu, \mathbf{q}}\vert^{2}\delta(\omega-\omega_{\mu, q}) \nonumber \\
 & = & \sum_{\mu \in \lbrace L, T \rbrace} \frac{V}{(2\pi)^{3}} \int \! \ud^{3}\mathbf{q} \vert\gamma_{\mu, \mathbf{q}}\vert^{2}\delta(\omega-\omega_{\mu, q}) \nonumber \\
 & = & J_{L}(\omega)+J_{T}(\omega)
\eea
where $\omega_{\mu, q} = c_{\mu}q$ and $J_{L}(\omega)$, $J_{T}(\omega)$ are the respective longitudinal and transverse acoustic phonon contributions to the spectral density.

\subsection{Longitudinal acoustic phonons}
For the contribution from longitudinal acoustic phonons, we find
\bea
J_{L}(\omega) & = & \frac{V}{(2\pi)^{3}}\int \! \ud^{3}\mathbf{q}\vert\gamma_{L, \mathbf{q}}\vert^{2}\delta(\omega-\omega_{L, q}) \nonumber \\
& = & \frac{\hbar\omega^{3}}{16 \pi^{3} \rho c_{L}^{5}} \int_{-1}^{1} \! \ud (\cos(\theta)) \int_{0}^{2\pi} \! \ud \varphi \Big[\Xi_{d}+\Xi_{u}\cos^{2}(\theta)\Big]^{2} \nonumber \\
& \times & \sin^{2}\Big( \frac{\omega}{\omega_{L,L}} \sin(\theta) \cos(\varphi) \Big) \exp \Big[ \frac{-\omega^{2}}{2 \omega_{L,a}^{2}} \sin^{2}(\theta) \Big] \nonumber \\
& \times & \exp \Big[ \frac{-\omega^{2}}{2 \omega_{L,b}^{2}} \cos^{2}(\theta) \Big] \nonumber \\
& = & \frac{\hbar \omega^{3}}{ 8 \pi^{2} \rho c_{L}^{5}} \exp \Big[ {\frac{-\omega^{2}}{2 \omega_{L,a}^{2}}} \Big] \int_{0}^{1} \! \ud v \Big[\Xi_{d} + \Xi_{u} v^{2} \Big]^{2} \label{eq:JLong} \\
& \times & \Big[1 - J_{0}\Big( \frac{2\omega \sqrt{1-v^{2}}}{\omega_{L,L}} \Big) \Big] \exp \Big[ \frac{-v^{2}}{2} \Big( \frac{\omega^{2}}{\omega_{L,b}^{2}} - \frac{\omega^{2}}{\omega_{L,a}^{2}} \Big) \Big], \nonumber
\eea
where $\omega_{L,L} = c_{L}/L$, $\omega_{L,a}=c_{L}/a$, $\omega_{L,b}=c_{L}/b$, and $J_{0}(x)$ is the zeroth Bessel function of the first kind. 
Equation (\ref{eq:JLong}) follows from the integral definition of the $n^{\mathrm{th}}$ Bessel function of the first kind
\be
J_{n}(x) = \frac{1}{\pi} \int_{0}^{\pi} \! \ud \theta \cos(n \theta - x \sin \theta).
\ee
Note that $J_{0}(x) \approx 1 - x^{2}/4 + \mathcal{O}(x^{4})$ for $\vert x \vert \ll 1$, so in the low energy regime, $\vert \omega / \omega_{L,L} \vert \ll 1$, the spectral density function scales as $J_{L}(\omega) \propto \omega^{5}$. 

\subsection{Transverse acoustic phonons}
Following the same analysis as above, for the contribution from transverse acoustic phonons we find
\bea
J_{T}(\omega) & = & \frac{V}{(2\pi)^{3}}\int \! \ud^{3}\mathbf{q}\vert\gamma_{T, \mathbf{q}}\vert^{2}\delta(\omega-\omega_{T,q}) \nonumber \\
& = & \frac{\Xi_{u}^{2} \hbar\omega^{3}}{16 \pi^{3} \rho c_{T}^{5}} \int_{-1}^{1} \! \ud (\cos(\theta)) \int_{0}^{2\pi} \! \ud \varphi \sin^{2}(\theta) \cos^{2}(\theta) \nonumber \\
& \times & \sin^{2}\Big( \frac{\omega}{\omega_{T,L}} \sin(\theta) \cos(\varphi) \Big) \exp \Big[ \frac{-\omega^{2}}{2 \omega_{T,a}^{2}} \sin^{2}(\theta) \Big] \nonumber \\
& \times & \exp \Big[ \frac{-\omega^{2}}{2 \omega_{T,b}^{2}} \cos^{2}(\theta) \Big] \nonumber \\
& = & \frac{\hbar \omega^{3}}{ 8 \pi^{2} \rho c_{T}^{5}} \exp \Big[ {\frac{-\omega^{2}}{2 \omega_{T,a}^{2}}} \Big] \int_{0}^{1} \! \ud v \Xi_{u}^{2} v^{2} (1-v^{2}) \label{eq:JTrans} \\
& \times & \Big[1 - J_{0}\Big( \frac{2\omega \sqrt{1-v^{2}}}{\omega_{T,L}} \Big) \Big] \exp \Big[ \frac{-v^{2}}{2} \Big( \frac{\omega^{2}}{\omega_{T,b}^{2}} - \frac{\omega^{2}}{\omega_{T,a}^{2}} \Big) \Big], \nonumber
\eea
where $\omega_{T,L} = c_{T}/L$, $\omega_{T,a}=c_{T}/a$, and $\omega_{T,b}=c_{T}/b$. 
As for the contribution from longitudinal acoustic phonons, $J_{T}(\omega) \propto \omega^{5}$ for $\vert \omega / \omega_{L,L} \vert \ll 1$.

\section{Error analysis}
In this experiment, we repeat the differential charge occupation measurement at each detuning and frequency value 100 times. 
The experimental error bars shown in Fig. 2 of the paper correspond to $\pm$ twice the standard error of the mean over these 100 measurements. 
We smooth the ensemble-averaged differential charge sensing data by evaluating a Fourier decomposition in terms of even (symmetric about $\epsilon=0$) Fourier modes. 
With an appropriate normalization, this enforces the physical constraint that $n(\epsilon) \in [0,1]$ and $n(0) = 0.5$, necessary for performing the subsequent fit. 
Making the assumption that the errors for each measured differential charge sensing value are independent and identically normally distributed, we derive error bars for the charge occupation $n(\epsilon,f)$. 
With information about the experimental noise statistics, we can produce as many ``noise'' realizations as we wish by adding appropriately normally-distributed noise to the smoothed and normalized mean values of the differential occupation. 
We define the misfit between two given occupation functions $n_{1}(\epsilon,f)$ and $n_{2}(\epsilon,f)$ as
\be
\mathcal{M} = \sum_{i,j} \vert n_{1}(\epsilon_{i}, f_{j}) - n_{2}(\epsilon_{i}, f_{j}) \vert^{2},
\ee
where $\epsilon_{i}, f_{j}$ are respectively the detuning and frequency values measured. 
We then evaluate an effective standard deviation of the squared-deviation misfit $\delta \mathcal{M}$ between the smoothed occupation data and the noise-added data. 
We optimize over splined representations of the relaxation rate $\Gamma(\epsilon)$, finding a best fit to the smoothed occupation data $\Gamma$ corresponding to the misfit $\mathcal{M}_{\mathrm{min}}$. 
Then we individually perturb each of the components of the vector parameterizing the spline for $\Gamma(\epsilon)$ until the misfit becomes $\mathcal{M}_{\mathrm{min}} + \delta \mathcal{M}$. 
This analysis defines the confidence regions for the relaxation rate, plotted in Fig. 4.

\bibliography{mybibliography}